\documentclass[amsmath,
11pt, 
amssymb,
nofootinbib,
notitlepage,
article,
superscriptaddress,
]{revtex4-1}
\usepackage[T1]{fontenc}
\usepackage[usenames,dvipsnames]{xcolor}
\usepackage{graphicx,hyperref,orcidlink,float}
\hypersetup{colorlinks=true,citecolor=RoyalBlue,linkcolor=RoyalBlue,urlcolor=RoyalBlue}
\usepackage{comment}
\usepackage{appendix}
\usepackage{graphicx}

\usepackage{soul}
\usepackage{tikz}
\usepackage{pgfplots}
\usepackage{dcolumn}
\usepackage{bm}
\usepackage{amsmath}

\usepackage[mathlines]{lineno}
\usepackage{xcolor}
\usepackage{slashed}

\usepackage{lineno}
\usepackage{xr}

\makeatletter
\newcommand*{\addFileDependency}[1]{
\typeout{(#1)}
\@addtofilelist{#1}
\IfFileExists{#1}{}{\typeout{No file #1.}}
}\makeatother

\newcommand*{\myexternaldocument}[1]{%
\externaldocument{#1}%
\addFileDependency{#1.tex}%
\addFileDependency{#1.aux}%
}
\myexternaldocument{nature-methods}

\usepackage[normalem]{ulem}

\newcommand{\stf}[1]{\langle #1 \rangle}

\pgfplotsset{compat=1.16}
\begin{document}

\title{A constraint on the dissipative tidal deformability of neutron stars}

\author{Justin L. Ripley}

\affiliation{Illinois Center for Advanced Studies of the Universe \& Department of Physics, University of Illinois at Urbana-Champaign, Urbana, IL 61801, USA}

\author{Abhishek Hegade K.R.}

\affiliation{Illinois Center for Advanced Studies of the Universe \& Department of Physics, University of Illinois at Urbana-Champaign, Urbana, IL 61801, USA}

\author{Rohit S. Chandramouli}

\affiliation{Illinois Center for Advanced Studies of the Universe \& Department of Physics, University of Illinois at Urbana-Champaign, Urbana, IL 61801, USA}

\author{Nicol\'as Yunes}

\affiliation{Illinois Center for Advanced Studies of the Universe \& Department of Physics, University of Illinois at Urbana-Champaign, Urbana, IL 61801, USA}

\date{\today}

\begin{abstract}
The gravitational waves (GWs) emitted by neutron star binaries probe the physics of matter at supra nuclear densities.
During the late inspiral, tidal deformations raised on each star by the gravitational field of its companion depend crucially on the star's internal properties.
The misalignment of a star's tidal bulge with its companion's gravitational field encodes the strength of internal dissipative processes, which imprint onto the phase of the gravitational waves emitted.
We here analyze GW data from the GW170817 (binary neutron star) event detected by LIGO and Virgo and find the first constraint on the dissipative tidal deformability of a neutron star.
From this constraint, \emph{assuming} a temperature profile for each star in the binary, we obtain an order of magnitude bound on the averaged bulk ($\zeta$) and shear ($\eta$) viscosity of each star during the inspiral.: $\zeta \lesssim 10^{31} \mathrm{g}\;\mathrm{cm}^{-1}\mathrm{s}^{-1}$ and $\eta \lesssim 10^{28} \mathrm{g}\;\mathrm{cm}^{-1}\mathrm{s}^{-1} $. 
We forecast that these bounds could be improved by two orders of magnitude with third-generation detectors, like Cosmic Explorer, using inspiral data.
These constraints already inform nuclear physics models and motivate further theoretical work to better understand the interplay between viscosity and temperature in the late inspiral of neutron stars.  
    
\end{abstract}

\maketitle

Neutron stars are the densest material objects in the universe, with their interior reaching densities many times that of atomic nuclei at temperatures well below those probed by heavy ion collisions. 
Determining the physical properties of neutron star matter, therefore, has remained an outstanding problem in the fields of astrophysics, gravitational physics, and nuclear physics for almost a century~\cite{Burgio:2021vgk}.
Gravitational waves (GWs) from neutron star binaries encode the tidal deformations that neutron stars experience during their late inspiral before merger.
These tidal deformations are in turn affected by the material properties of the stars. 
The prompt, conservative, relativistic tidal response is described by the tidal deformability $\Lambda$ \cite{Hinderer:2007mb,Damour:2009vw,Binnington:2009bb}, which encodes aspects of the \emph{equilibrium} properties of neutron star matter--the neutron star equation of state (EOS) \cite{Burgio:2021vgk}.~The GWs from the binary neutron star event GW170817 constrained the tidal deformability of neutron stars, and thus, and thus the EOS of those stars~\cite{LIGOScientific:2017vwq,LIGOScientific:2018hze}. 

\begin{figure}[ht!]
    \centering
    \includegraphics[width = 0.45\textwidth]{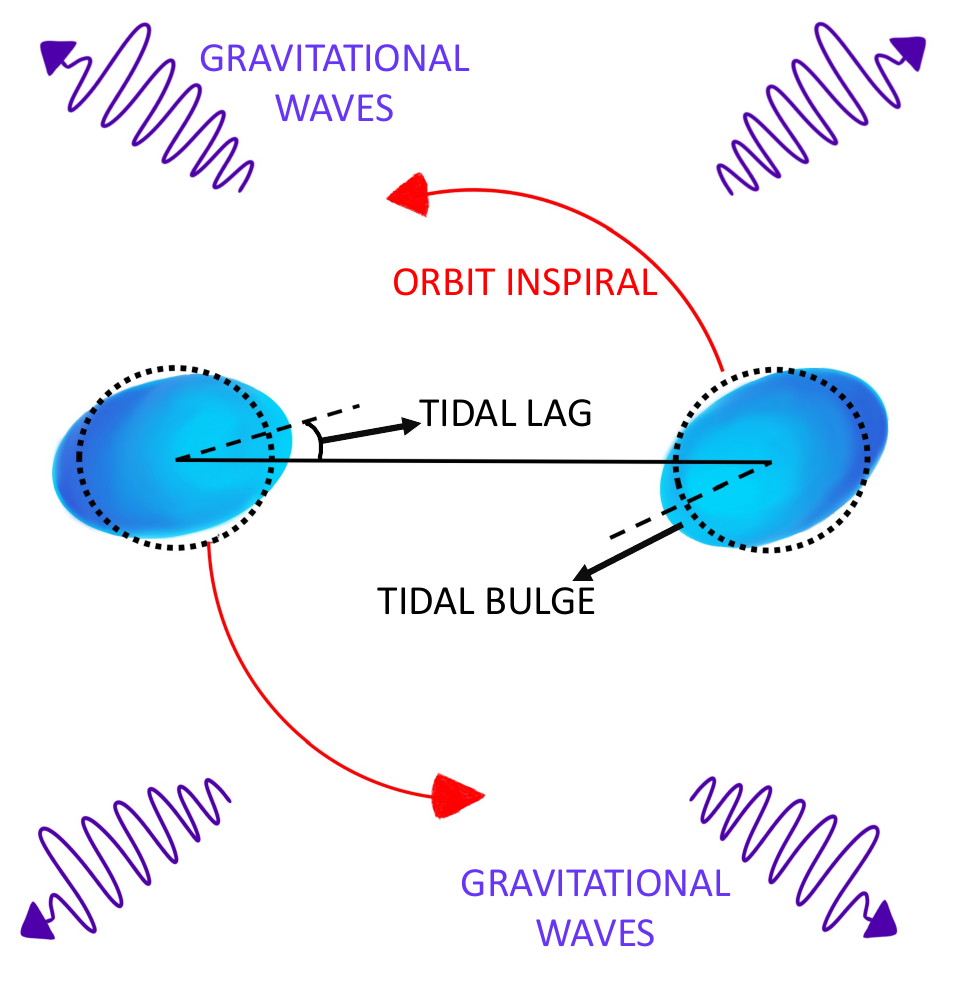}
    \caption{Cartoon of the tidal responses of two stars in a quasi-circular binary (not to scale).
    Dissipative, out-of-equilibrium effects force the tidal bulge of each star to be misaligned with the gravitational field sourced by its companion. 
    }
    \label{fig:illustration-tidal-lag-binary}
\end{figure}

The tidal response of a star can be visualized as a ``tidal bulge'' that is sourced but not aligned with the time-varying, externally-imposed gravitational field of its companion (see figure~\ref{fig:illustration-tidal-lag-binary}). 
Non-equilibrium, dissipative effects within the star force the bulge to trail the orbit, inducing a tidal lag angle between the direction of the bulge and the orbital separation. 
The extent to which the tidal multipolar moments are misaligned with the external gravitational multipolar moments is, to a first approximation, described by the tidal lag time $\tau_d$. 
The tidal lag time is a universal feature for self-gravitating astrophysical objects, from planets \cite{Ogilvie_2014} to black holes \cite{Hartle:1973zz,Poisson:2009di}, and it has been observed in many planetary systems~\cite{Zahn:2008fk,Ogilvie_2014}. 
This tidal misalignment torques the two stars and heats them up through tidal, viscous heating (e.g. \cite{Lai:1993di,Arras:2018fxj}). 

The magnitude of these effects on the dynamics of neutron star binaries had been thought since the 1990s to be too small to measure with gravitational waves. 
Indeed, unphysically large values of viscosity are required to tidally lock the spins of the two stars to their orbit before merger~\cite{1992ApJ...400..175B}.
Recent work in nuclear physics, however, suggests that weak-force processes can induce an effective bulk viscosity~\cite{Alford:2019qtm,Most:2021zvc,Yang:2023ogo}, which, although not large enough to tidally lock the spins of the two stars, may still have a measurable effect in the gravitational waves emitted during the late inspiral~\cite{Ripley:2023qxo}. 
This has motivated different numerical groups to model out-of-equilibrium effects during the late inspiral, merger and the post-merger phase. 
Some groups have found bulk viscous effects to be enhanced during the late inspiral~\cite{Most:2021zvc,chabanov2023impact}. 
Meanwhile, other groups working with moment-based treatments of neutrino transport have not found evidence for large out-of thermodynamic equilibrium effects necessary for producing an effective bulk viscosity during the late inspiral~\cite{Radice:2021jtw}. 
Nevertheless, these  moment-based treatments of neutrino transport have revealed evidence of bulk viscous effects within a small window after the merger before matter goes back to equilibrium~\cite{espino2023neutrino}.
Given these differences in the literature, it is crucial to utilize the available gravitational wave data to provide insights into out-of-equilibrium effects present during the late inspiral and merger phase.

Recently, the signature of the tidal lag in the gravitational waves emitted by binary neutron star inspirals was re-analyzed in \cite{Ripley:2023qxo}, and found to be parametrically enhanced relative to that of conservative tidal effects. 
Conservative tidal effects first enter the GW phase proportional to ten powers of the orbital velocity relative to the leading-order term in a small-velocity, post-Newtonian (PN) expansion, which in the late inspiral is $v \sim (0.25$--$0.4)c$; moreover, conservative effects are inversely proportional to five powers of the stellar compactness, i.e., the dimensionless ratio of its mass to its radius, which for neutron stars is $C \sim 0.1$--$0.3$~\cite{Flanagan:2007ix,Hinderer:2007mb}.
In contrast, dissipative tidal effects first enter the GW phase proportional to eight powers of the orbital velocity and inversely proportional to six powers of the stellar compactness~\cite{Ripley:2023qxo}. 
Therefore, this parametric enhancement boosts the effect of dissipative tidal effects, making them potentially measurable with \emph{current} ground-based GW detectors for physically plausible levels of dissipation within the stars.

The contribution of the tidal lag to the GW phase is captured by the dissipative counterpart to the conservative tidal deformability $\Lambda$:  the dissipative tidal deformability $\Xi$, which can be mapped to the effective bulk/shear viscosities of the star \cite{Ripley:2023qxo,HegadeKR:2024agt}. 
Large values of bulk viscous dissipation could be sourced through Urca processes~\cite{physrevd.39.3804,Alford:2019qtm,Most:2021zvc,Yang:2023ogo}, or more exotic nuclear processes due to the presence of hyperons deep in the interior of a neutron star \cite{Jones:2001ya,Lindblom:2001hd,Alford:2020pld,Gusakov:2008hv}. 
This effective viscosity depends on the particular nuclear and fluid models used, on the EOS of the star, and on its temperature. 
Therefore, a measurement or constraint on $\Xi$ cannot be translated to an independent measurement or constraint of any single one of these ingredients, without an assumption about the values of any two of the other quantities.   
Nonetheless, taking the expected values for these quantities, a preliminary forecast in \cite{Ripley:2023qxo} suggested that the total extent of dissipative processes within neutron stars could be meaningfully constrained through an analysis of current GW data. 
We here carry out precisely such an analysis and constrain the dissipative tidal deformability of neutron stars using data from the GW170817 event detected by LIGO and Virgo~\cite{LIGOScientific:2017vwq}.
\section{Gravitational-wave model with dissipative tidal effects}
We use the \texttt{IMRPhenomPv2\_NRTidal} waveform model \cite{Dietrich:2019kaq} for the detector response to the impinging gravitational waves emitted in the inspiral of binary neutron star system, which we enhance to include tidal dissipation. 
Without tidal dissipation, the \texttt{IMRPhenomPv2\_NRTidal} model depends on $17$ parameters $\theta_{a}$.
These parameters, in addition to the non-tidal ones (such as the chirp mass ${\cal{M}} \equiv m_{A}^{3/5} m_{B}^{3/5}/M^{1/5}$ and the symmetric mass ratio $\eta_{\mathrm{sym}} \equiv m_{A} m_{B}/M^{2}$, with $M \equiv m_{A}+m_{B}$ the total mass and $m_{A/B}$ the component masses), includes the tidal deformabilities of the stars $\Lambda_{A/B}$.
In the frequency domain, we represent the model via the Fourier transform of the GW strain $\tilde{h}\left(f\right)= A\left(f;\theta\right) e^{i\Psi\left(f;\theta\right)}$, where $A(f;\theta)$ is the Fourier GW amplitude and $\Psi(f;\theta)$ is the Fourier GW phase. 

We enhance the \texttt{IMRPhenomPv2\_NRTidal} model by adding the leading PN contribution of the dissipative tidal deformability to the \texttt{IMRPhenomPv2\_NRTidal} Fourier phase \cite{Ripley:2023qxo}
\begin{align}
\label{eq:dissipative-tidal-phase-contribution}
    \Psi(f;\theta) = \Psi_{{\rm Pv2NRT}}(f;\theta_{a}) -
    \frac{225}{4096}\frac{1}{\eta_{\mathrm{sym}}}\tilde{\Xi}\;u^3\log\left(u\right)
    ,
\end{align}
where $u\equiv \left(G\pi M f/c^3\right)^{1/3}$ is effectively the orbital velocity. 
Since the leading, PN order term in $\Psi_{{\rm Pv2NRT}}$ is proportional to $u^{-5}$, we see that the dissipative tidal contribution is of ${\cal{O}}(u^{8})$ relative to leading order, which is of ${\cal{O}}(u^{2})$ {\emph{larger}} than the ${\cal{O}}\left(u^{10}\right)$ relative conservative tidal contribution in $\Psi_{{\rm Pv2NRT}}$. 
The quantity $\tilde{\Xi}$ is the binary dissipative ``chirp'' tidal deformability, which is a weighted sum of the dissipative tidal deformabilities of each star, $\Xi_{A,B}$
\begin{align}
    \label{eq:combined-dissipative-tidal-deformability}
    \tilde{\Xi}
    \equiv \,&
    8\left(2\eta_{\mathrm{sym}}^2 - 4\eta_{\mathrm{sym}} + 1\right)
    \Xi_s
    -
    8\sqrt{1-4\eta_{\mathrm{sym}}}\left(1-2\eta_{\mathrm{sym}}\right)
    \Xi_a
    ,
\end{align}
where $\Xi_s\equiv\left(\Xi_A+\Xi_B\right)/2$ and $\Xi_a\equiv\left(\Xi_B-\Xi_A\right)/2$. Therefore, the enhanced \texttt{IMRPhenomPv2\_NRTidal} model contains $18$ parameters: $\theta = \theta_{a} \cup \{\tilde{\Xi}\}$. 

\section{Constraint on dissipative tidal deformability from GW170817}
As we discuss in more detail in Methods, we use a Bayesian parameter estimation to compute the posterior probability distribution for all the parameters of our enhanced \texttt{IMRPhenomPv2\_NRTidal} GW model, given the publicly available 4kHz GW170817 GW strain data \cite{LIGOScientific:2019lzm}. 
While we performed several checks of our analysis, here we describe only two separate analysis (we describe our other checks in Methods). 
In one run, we sampled in the stars' individual conservative tidal deformabilities $\Lambda_{A,B}$, and in the other we sampled on the symmetric tidal deformability $\Lambda_s = (\Lambda_A + \Lambda_B)/2$, from which we find $\Lambda_a = (\Lambda_B - \Lambda_A)/2$ through the binary Love relations \cite{Yagi:2015pkc} (we marginalize over the uncertainty in those relations following~\cite{Carson:2019rjx}).

We present the binned, marginalized posterior distribution on $\tilde{\Xi}$ from those two analyses in figure~\ref{fig:marginalized-xibar-GW170817}. 
We see that the data is informative, yielding a posterior that is significantly different than the prior and peaked at zero with $\tilde{\Xi}<1200$ at 90\% confidence. The posterior is additionally independent of whether or not we use the marginalized binary Love relations. 

\begin{figure}[ht]
   \centering
   \includegraphics[width=0.6\columnwidth]{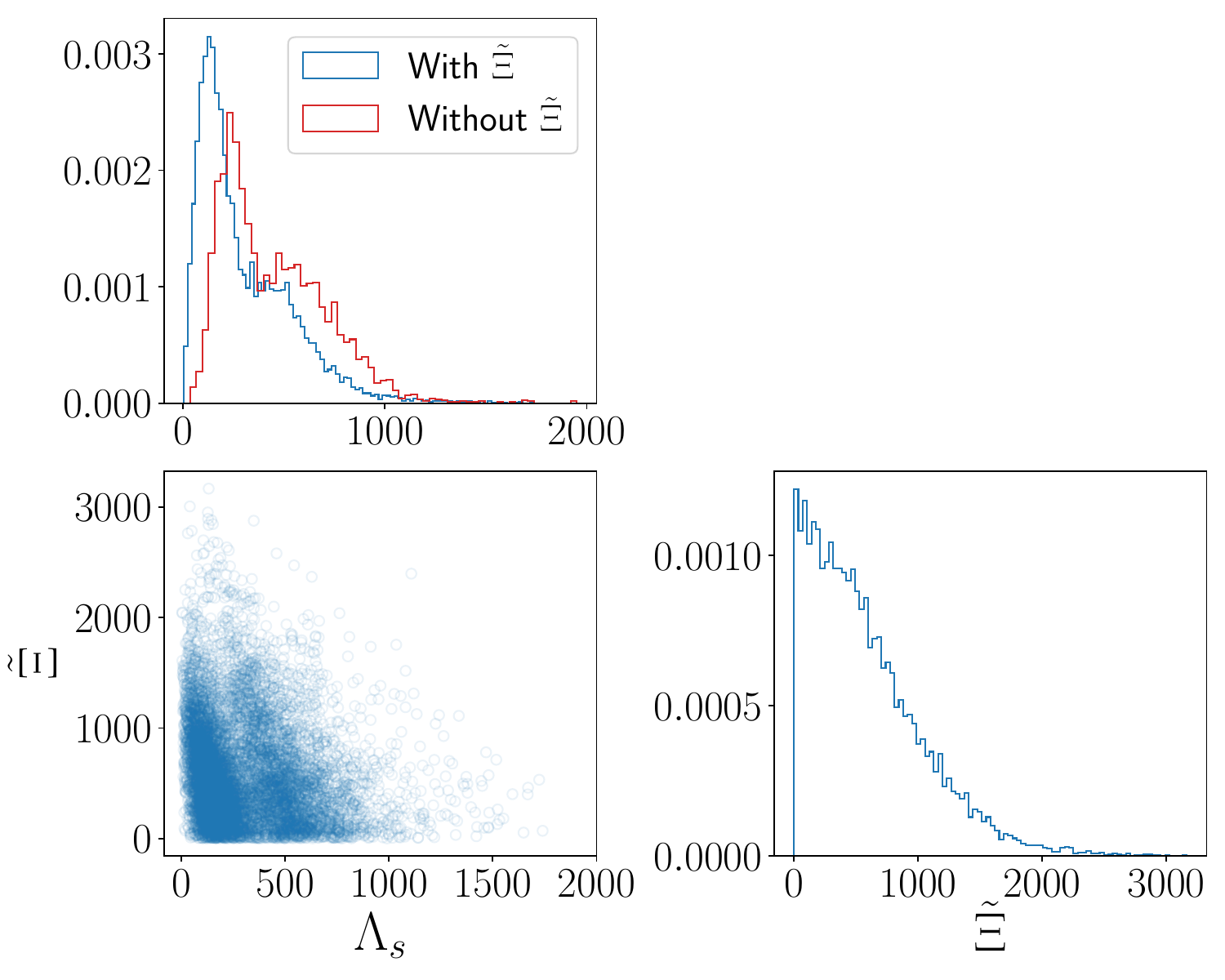}
   \caption{
   Corner subspace of the marginalized posterior distribution of $\Lambda_s$ and $\tilde{\Xi}$.
   Shown in blue are the 1-dimensional and 2-dimensional marginalized posteriors on $\Lambda_s$ and $\tilde{\Xi}$ corresponding to the analysis case where we do not use binary love relations. 
   Additionally shown in red in the top left panel is the 1-dimensional marginalized posterior of $\Lambda_s$ from the LIGO--Virgo analysis where $\tilde{\Xi}$ was not included (without binary love relations). 
   Observe that the two parameters are correlated.
   This correlation pushes the 1-d marginalized distribution of $\Lambda_s$ to slightly lower values when including $\tilde{\Xi}$ in the analysis, as compared to when we exclude the $\tilde{\Xi}$ correction (as is done in the LIGO--Virgo analysis of GW170817 \cite{LIGOScientific:2018hze}).
}
   \label{fig:corner_lambdas_xibar}
\end{figure}

The marginalized posterior distributions for almost all other parameters are statistically consistent with those obtained by the LIGO-Virgo collaboration when one does not include dissipative tidal effects. The one exception is the posterior for $\tilde{\Lambda}$ (or equivalently for $\Lambda_{s}$ or $\Lambda_{A,B}$), which are pushed to lower values due to correlations between the conservative and dissipative tidal deformabilities, as shown in figure~\ref{fig:corner_lambdas_xibar}. 
This implies that if the dissipative tidal deformability is present in the signal, and one neglects to incorporate its effects in the waveform model, one will then be biased in the estimation of the conservative tidal deformability to higher values than those contained in the signal. 

\section{Implications for nuclear physics}

Any dissipative process within a neutron star adds cumulatively to the dissipative tidal deformability.
As a consequence of this, an upper bound on $\tilde{\Xi}$ constrains the strength of any given dissipative process.
Given a fluid model for the star, we can relate the dissipative tidal deformability to the tidal lag inside star $A$ via \cite{Ripley:2023qxo} 
\begin{align}
    \label{eq:parameterization_xiA}
    \Xi_{A}
    =
    \frac{2}{3} k_{2,A} \left(\frac{1}{C_{A}^6}\right)
    \left(\frac{c \tau_{d,A}}{R_{A}}\right)
    ,
\end{align}
where $k_{2,A}$ is the conservative tidal Love number of the star, $m_A$ and $R_{A}$ are the stellar mass and radius,
$C_A\equiv Gm_A/(R_Ac^2)$ is its compactness, and $\tau_{d,A}$ is its tidal lag time. 
We relate the tidal lag time to the viscosity via $\tau_{d,A}=p_{2,A} \stf{\delta}R_Ac^2/(Gm_A\stf{e})$ \cite{Poisson:2009di,Ripley:2023qxo}, 
where $\stf{\delta} = \stf{\eta} $ (the shear viscosity) or $ \stf{\zeta}$ (the bulk viscosity) depending on which source of dissipation is dominant;
$p_{2,A}$ is a dimensionless constant that is computed for a given viscosity profile $\delta(e)$, and $\stf{e}$ is the average energy density.
Using that the conservative tidal deformability $\Lambda_A\equiv (2k_{2,A}/3) / C_A^5$ \cite{Flanagan:2007ix}, we then obtain 
\begin{align}
    \label{eq:relation-xi-and-nu}
    \Xi_A
    &=
    \frac{c^3}{G} \frac{p_{2,A} \Lambda_A}{C_A} \frac{\stf{\delta}}{\stf{e} m_A}\,,
    \nonumber \\
    &\approx  26.1  \times  \left(\frac{p_{2,A}}{0.01}\right) 
    \left(\frac{\Lambda_A}{300}\right)  \left(\frac{0.188}{C_A}\right)  
     \left( \frac{\stf{\delta}}{10^{{30}} \frac{\rm g}{\rm cm s}} \right)
      \left( \frac{1.38 M_{\odot}}{m_A}\right)
       \left( \frac{9 \times 10^{34} \frac{\rm erg}{\rm cm^{3}}}{\stf{e}}\right).
\end{align}
When bulk viscosity drives the process of dissipation $p_{2,A}\sim 0.01$, but it can be as high as $p_{2,A} \sim 5$ when shear viscosity dominates~\cite{HegadeKR:2024agt}.
This is because
gravitational fields cause a larger relative shearing motion than compression of the star \cite{HegadeKR:2024agt}.

We can map the constraint we obtained on the dissipative tidal deformability to the microphysics of dissipative processes inside a neutron star if we make the following assumptions. We assume that both stars in the binary had the same EOS and the same mass (which is consistent with the posterior for the GW170817 event). 
If so, they must also have the same compactness $C_{A} = C_{B}$.
Furthermore we assume the stars have the same temperature profile and EOS, so that $\Lambda_{A} = \Lambda_{B} = \tilde{\Lambda}$ and $\Xi_A=\Xi_B=\tilde{\Xi}$. 
Inverting equation~\eqref{eq:relation-xi-and-nu}, we obtain
\begin{align}
    \label{eq:relating-zeta-and-xibar}
    \stf{\delta}_A
    \approx&\;
    4.6 \times 10^{31}\frac{\mathrm{g}}{\mathrm{cm}\;\mathrm{s}}
    \left(\frac{\tilde{\Xi}}{1200}\right)
    \left(\frac{\stf{e}}{9\times 10^{34} \mathrm{erg}\;\mathrm{cm}^{-3}}\right)
    \left(\frac{300}{\Lambda_A}\right)
    \left( \frac{0.01}{p_{2,A}}\right)
    \left(\frac{C_A}{0.188} \right)\left(\frac{m_A}{1.38 M_{\odot}}\right)
    .
\end{align}
If shear viscosity was the dominant contribution to the dissipation, we would then obtain $\stf{\eta} \approx 9.15 \times 10^{28} \mathrm{g}\;\mathrm{cm}^{-1}\mathrm{s}^{-1}$ with a normalization factor of $\left(5/p_{2,A}\right)$ instead of $\left(0.01/p_{2,A}\right)$. 
Given our measured bound of $\tilde{\Xi}\lesssim 1200$, we can place an upper bound on the averaged bulk and shear viscosity during the evolution of the GW170817 event.
We respectively obtain $\stf{\zeta}_A \lesssim 4.57 \times 10^{31}\mathrm{g}\;\mathrm{cm}^{-1}\mathrm{s}^{-1} $ and $\stf{\eta}_A \lesssim 9.15 \times 10^{28}\mathrm{g}\;\mathrm{cm}^{-1}\mathrm{s}^{-1}$.

We now put this constraint in context by comparing it to current theoretical estimates of the viscosity of neutron stars.
Viscosity generated by microscopic processes in neutron stars depends sensitively on the local stellar temperature profile $T$. 
Shear viscosity in neutron star cores scales as $T^{-2}$ due to electron-muon scattering~\cite{Shternin:2008es} and is expected to be less than $\stf{\eta}\lesssim 10^{22} \mathrm{g}\;\mathrm{cm}^{-1}\mathrm{s}^{-1}$.
Shear viscous contributions due to the interface of the stars crust with its interior have been speculated to be as large as $\stf{\eta}\sim 10^{29}\mathrm{g}\;\mathrm{cm}^{-1}\mathrm{s}^{-1}$ \cite{1992ApJ...398..234K}. 
Bulk viscous contributions due to the presence of hyperons scale as $T^{-2}$ and may be the dominant source of dissipation in neutron stars at very low ($\sim$keV) temperatures, with values of bulk viscosity predicted to exceed $\stf{\zeta}\sim 10^{30}\mathrm{g}\;\mathrm{cm}^{-1}\mathrm{s}^{-1}$ in some models \cite{Jones:2001ya,Lindblom:2001hd,Gusakov:2008hv,Alford:2020pld}. 
As the binary enters the late inspiral, heating from tidal friction due to Urca reactions may increase the temperature of the two stars to tens of keV \cite{Arras:2018fxj}; hyperonic bulk viscous contributions may heat the stars to higher temperatures \cite{Alford:2020pld}. 
Numerical relativity simulations of neutron star mergers suggest tidal heating could increase the stellar temperature to a few MeV during the last few orbits \cite{Perego:2019adq}. 
Bulk viscous contributions from direct and modified Urca processes are expected to dominate at high temperatures as those reactions scale as $T^{4}$ and $T^6$, respectively~\cite{physrevd.39.3804}. We note though that beyond a resonant peak, these reactions become weaker at higher temperatures; current estimates place this peak at $T\sim 5\mathrm{MeV}$ \cite{Alford:2023gxq}, higher than expected temperatures reached by the stars during the inspiral.
Typical predictions for the bulk viscosity for Urca-process-driven viscosity range from $\stf{\zeta}\sim 10^{26}\;\mathrm{g}\;\mathrm{cm}^{-1}\mathrm{s}^{-1}$ when $T\sim0.1\mathrm{MeV}$, to $\stf{\zeta}\sim 10^{31}\;\mathrm{g}\;\mathrm{cm}^{-1}\mathrm{s}^{-1}$ when $T\sim 1 \, \mathrm{MeV}$, depending on the EOS \cite{Most:2021zvc,Yang:2023ogo}.

In Methods, we perform an estimate of the number of GW cycles that $\tilde{\Xi}$ introduces in different regimes of the inspiral. 
We estimate that $\tilde{\Xi}\sim1200$ introduces $\sim 1.86$ GW cycles over the last $\sim 27$ orbits (corresponding roughly to orbital separations from $\sim 83$km to contact), and a total of $\sim 2.5$ GW cycles over the entire inspiral.
These results suggest that accounting for finite temperature and out-of-equilibrium effects during the last stage of the inspiral will be critical for mapping a constraint/detection of $\tilde{\Xi}$ to the underlying nuclear physics. 

\section{Future detectability of dissipative tides}
We simulate synthetic GW detections for three different networks: ground-based detectors in the O4 and O5 era (i.e. the current LIGO observation run setup, and the following upgrade with the addition of the LIGO-India detector) \cite{KAGRA:2013rdx}, and the Cosmic Explorer (CE) era (one CE instrument at the current Hanford detector site) \cite{LIGOScientific:2016wof}.
Each simulation consists of $128$s of synthetic data, all starting at $40$Hz; 
The component masses and the symmetric conservative tidal deformabilities in each simulation are kept fixed at $m_{A,B}=1.38 M_{\odot}$ and $\Lambda_s=584$ respectively, while the injected chirp dissipative tidal deformabilities are chosen to be  $\tilde{\Xi}=\{20,400,800,1200\}$. 
The $m_{A,B}$ and $\Lambda_{s}$ choices are consistent with the marginalized posterior distribution of these parameters in the GW170817 event, and the $\tilde{\Xi}$ choice is consistent with the constraint we obtained at 90\% confidence. 
Since the luminosity distance is kept fixed, the signal-to-noise ratio (SNR) increases in the O4, O5, and CE simulations to $\sim 60$, $\sim 100$, and $\sim 1100$, respectively.

We analyze these injections using Bayesian parameter estimation with our enhanced \texttt{IMRPhenomPv2\_NRTidal} waveform model, following almost the same data analysis procedure as for the GW170817 event. The only difference is that this time, we only sample on $\{\mathcal{M},q,\Lambda_s,\tilde{\Xi}\}$, and thus, we fix all other parameters in the posterior to their injected values; we find empirically that this does not qualitatively affect our conclusions. 
We emphasize that we use the same Uniform prior for $\tilde{\Xi}$ that we use in our analysis of GW170817; that is we do not use our marginalized posterior for $\tilde{\Xi}_A$ from GW170817 as our prior for $\tilde{\Xi}$ in our injection runs. 

\begin{figure}[h!]
    \centering
    \includegraphics[width =0.6\columnwidth]{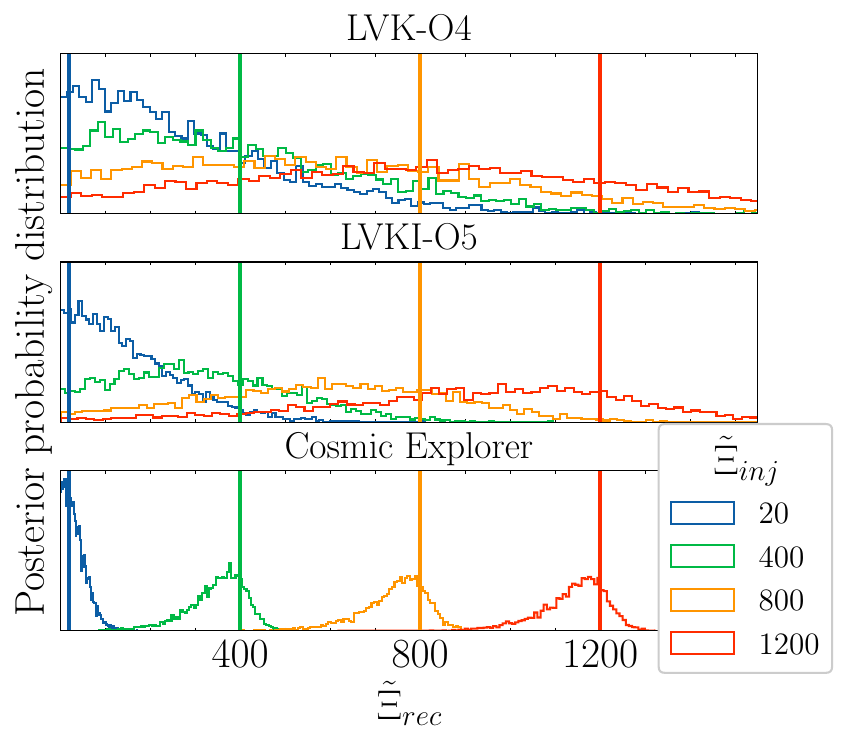}
    \caption{Histograms of the marginalized posterior distributions for $\tilde{\Xi}$ for zero-noise realizations of a GW170817-like event, simulated to be detected with the O$4$ LIGO-Virgo-KAGRA, O$5$ LIGO-Virgo-KAGRA-India, and Cosmic-Explorer detector networks.
    The injected value of $\tilde{\Xi}$ is denoted by $\tilde{\Xi}_{inj}$ and the recovered value of $\tilde{\Xi}$ is denoted by $\tilde{\Xi}_{rec}$. 
    The vertical lines represent the injected values of $\tilde{\Xi}$. 
    We use the same uniform prior for $\tilde{\Xi}$ as described in the text.}
    \label{fig:posterior-xibar-histogram}
\end{figure}

Figure~\ref{fig:posterior-xibar-histogram} shows the marginalized posterior distribution for $\tilde{\Xi}$ for each injection.
Notice that $\tilde{\Xi}$ is biased towards zero; we attribute this to correlations between $\tilde{\Xi}$ and the tidal deformability parameters $\Lambda_{A/B}$ (see Methods for more details).
The one exception to this bias is the $\tilde{\Xi}_{inj}=20$ injection for the O$4$ network; we attribute this increase to the fact that the spread in the posterior is very large for that network, and the injected value is close to zero (the lower bound of our prior).
As expected, given the values of the SNR for the three detector networks, from figure~\eqref{fig:posterior-xibar-histogram} we see that the O5 network gives a modest improvement to the measurement of $\tilde{\Xi}$.
For the O4 network, we see that the posterior is mostly supported away from zero only for the largest injected value of $\tilde{\Xi}=1200$, which lies at the $90\%$ credible interval for our current measurement from GW170817 data.
Table~\eqref{fig:posterior-xibar-table}, shows the maximum of the posterior distribution (\emph{maximum-a-priori} estimate, or MAP) for the marginalized distributions for $\tilde{\Xi}_{rec}$, along with the 90\% symmetric credible interval about the MAP.
With increasingly sensitive detectors (or alternatively, with increasingly high SNR), the measurement of $\tilde{\Xi}$ simultaneously becomes less biased and more precise.
This is consistent with our interpretation of the bias of $\tilde{\Xi}$ as arising from the correlation between $\tilde{\Xi}$ and $\Lambda_{A,B}$ -- with increasing SNR, these parameters simultaneously become less biased and can be measured more precisely. 
\section{Conclusions and outlook} 

Our analysis opens the door to a plethora of future work that will be required to extract precise inferences about out-of-equilibrium microphysics. 
First, one must find a way to break the degeneracy between the individual dissipative tidal deformabilities $\Xi_{A,B}$, which enter the GW phase through a linear combination encapsulated in $\tilde{\Xi}$. 
For the conservative tidal deformability $\tilde{\Lambda}$, this can be done through the binary Love relations~\cite{Yagi:2015pkc,Yagi:2016qmr} or through the measurement of higher PN order terms in the phase~\cite{Vines:2011ud}. 
One should thus investigate the existence of similar, approximately-universal relations for the dissipative tidal deformabilities, and calculate the higher-order PN dissipative terms. 
Second, the effective bulk and shear viscosities of neutron stars are predicted to depend somewhat on the stars EOS (e.g. \cite{physrevd.39.3804,Haensel:2000vz,Lindblom:2001hd,Jones:2001ya,Yang:2023ogo,Shternin:2008es}). 
Mapping inferences on $\Xi_{A,B}$ to micro-physics will thus require that one marginalize over all equations of state consistent with data, including those with sharp features in the speed of sound (e.g.~\cite{Tan:2021nat,Tan:2021ahl}). 
Third, it will be critical to investigate systematic biases in parameter estimation that arise when not including the dissipative tidal effects, and statistical biases in the measurement of $\tilde{\Xi}$ that arise due to the late inspiral being buried under detector noise.
Finally, the bulk and shear viscosity of neutron stars also depends sensitively on their local temperature. 
Connecting the individual tidal deformabilities $\Xi_{A,B}$ to nuclear physics, therefore, requires knowledge of the temperature profiles within each star, which generally increase with time (and thus with GW frequency) in the late inspiral due to tidal friction \cite{Lai:1993di,Arras:2018fxj}. 
This opens the possibility of learning not just about viscosity \emph{per se}, but also about the local temperature evolution and new aspects of the EOS of neutron stars.

\section{Methods\label{sec:methods}}
\subsection{Data analysis methodology\label{sec:data-analysis-robustness}}
We use a Bayesian parameter estimation to compute the posterior probability distribution for all the parameters of our enhanced \texttt{IMRPhenomPv2\_NRTidal} GW model, given 128s of the publicly available 4kHz GW170817 (glitch-cleaned) GW strain data \cite{LIGOScientific:2019lzm}. 
We sample over all $18$ parameters of the model and we marginalize over the reference phase.
As usual in GW data analysis, we assume the noise to be Gaussian and stationary; the log-likelihood of the strain data $\tilde{s}(f)$ given a GW template $\tilde{h}(f;\theta)$ with model parameters $\theta$ is then~\cite{Maggiore-vol-1} 
\begin{align}
    \ln \mathcal{L}  (\tilde{s}| \theta)
    &= - \dfrac{1}{2}\left(
                \tilde{r}(\theta)|
                \tilde{r}(\theta)
            \right) 
    =
    -
    2 \int_0^{\infty} df 
        \frac{
        \left|\tilde{r}(f;\theta)\right|^2          
        }{
            S_n\left(f\right)
        }
        ,
\end{align}
where $\tilde{r}(f;\theta) \equiv \tilde{h}\left(f;\theta\right) - \tilde{s}\left(f\right) $ is the residual signal and $S_n\left(f\right)$ is the noise power spectral density of the GW detector.

We use the \texttt{Bilby} \cite{Ashton:2018jfp} GW library, modified to incorporate the dissipative tidal effects of equation~\eqref{eq:dissipative-tidal-phase-contribution}, and sample the likelihood with a nested sampling algorithm as implemented in \texttt{DYNESTY} \cite{2020MNRAS.493.3132S}. 
Within the \texttt{Bilby} interface to that code, we set \texttt{nlive}$=1500$, \texttt{nact}$=10$, \texttt{dlogz}$=0.01$, \texttt{sample}$=$`\texttt{rwalk}', and \texttt{bound}$=$`\texttt{live}'.
As a consistency check on the convergence of our sampler to the true posterior distribution, we considered runs where \texttt{nlive}=1000, \texttt{nact}=5, and \texttt{dlogz}$=0.1$, and found that the posterior distribution did not significantly change.

We choose the following priors for all our parameter estimation analysis. 
We set the chirp mass $\mathcal{M}$ to lie within the range $\left[1.184\mathrm{M}_{\odot},1.25\mathrm{M}_{\odot}\right]$, with a distribution that is equivalent to two uniform distributions over the component masses $M_{A,B}$.
The mass ratio $q$ lies within the range $\left[0.5,1\right]$, and is also sampled uniformly in the two component masses.
As the \texttt{IMRPhenomPv2\_NRTidal} model does not include spin corrections to the conservative tidal effects, we use the ``low-spin'' prior defined in \cite{LIGOScientific:2018hze}; that is we use uniform priors for the neutron star spins $a_A,a_B$ over the range $[0,0.05]$. 
We note that our dissipative tidal term contains no spin corrections either.
When we sample on the tidal deformabilities separately, we use a uniform prior over the interval $\left[0,3000\right]$ for both deformabilities.
When we sample on the symmetric tidal deformability, we use a triangular prior with mean $1500$ and range $\left[0,3000\right]$.
For the dissipative tidal deformability $\tilde{\Xi}$,  we choose a uniform prior in $[0,8000]$. 
Our lower prior on $\tilde{\Xi}$ is set to zero because we exclude the possibility of anti-dissipative processes within each star ($\Xi_{A}<0$).
The upper end of the prior on $\tilde{\Xi}$ is set by a heuristic constraint on the timescale for causal momentum transport across the star: dissipative/viscous effects should not transport momentum faster than light speed \cite{1992ApJ...400..175B,Ripley:2023qxo}. 
The rest of our waveform parameter priors follow those of \cite{LIGOScientific:2018hze}.

\subsection{Validating our model and statistical analysis}

To check our data analysis methodology, we sampled both independently in $\Lambda_{A,B}$, and by sampling the symmetric tidal deformability $\Lambda_s = (\Lambda_A + \Lambda_B)/2$, from which we find $\Lambda_a = (\Lambda_B - \Lambda_A)/2$ through the binary Love relations \cite{Yagi:2015pkc}.
In the latter case, we marginalize over the uncertainty in the binary Love relations, following~\cite{Carson:2019rjx}.
To quantify possible systematic sources of error from using a particular base waveform model, we considered two additional analyses where we added the correction equation~\eqref{eq:dissipative-tidal-phase-contribution} to two different waveform models, \texttt{IMRPhenomPv2\_NRTidal} \cite{Dietrich:2018uni} and \texttt{IMRPhenomD\_NRTidal} \cite{Dietrich:2017aum}.
We found that the differences in the predictions for $\tilde{\Xi}$ for all three of these models (\texttt{IMRPhnomPv2\_NRTidal} with with and without the binary Love relations, and \texttt{IMRPhenomD\_NRTidal} with the binary Love relations) was insignificant for GW170817 strain data. 

We performed an additional check on our methods by performing the same parameter estimation procedure without $\tilde{\Xi}$ 
(i.e. by just using \texttt{IMRPhenomPv2\_NRTidal}), and confirming that our results were statistically consistent with the LIGO--Virgo analysis \cite{LIGOScientific:2018hze} of GW170817. 
We note that when we include $\tilde{\Xi}$ in our analysis, the marginalized posterior on $\Lambda_s$ is pushed to slightly lower values compared to when $\tilde{\Xi}$ is excluded. 
We quantify the bias in the measurement of $\Lambda_s$ by computing the difference between the maximum likelihood (ML) values of $\Lambda_s$ obtained with and without $\tilde{\Xi}$. 
We find that $\Delta \Lambda_{s}^{\mathrm{ML}} = |\Lambda^{\mathrm{ML, with} \ \tilde{\Xi}}_s - \Lambda^{\mathrm{ML, without} \ \tilde{\Xi}}_s| \approx 46.78$. 
The half-width of the 90\% credible interval of the marginalized distribution of $\Lambda_s$ in each case are $ \delta_{\Lambda_s}^{\mathrm{with} \ \tilde{\Xi}} \approx 291.49$ and $ \delta_{\Lambda_s}^{\mathrm{without} \ \tilde{\Xi}} \approx 381.00$. 
To conclude: the bias in maximum likelihood of $\Lambda_s$ is contained within both estimates of the statistical error.  

We see from the bottom left panel of figure~\eqref{fig:corner_lambdas_xibar} that the correlation between $\Lambda_s$ and $\tilde{\Xi}$ is the reason for the shift of the marginalized posterior on $\Lambda_s$.
That is, our agnostic analysis performed by including both conservative and dissipative tides results in slightly tighter constraints on the maximum value of the conservative tides $\Lambda_A$ and $\Lambda_B$ than is obtained by the LIGO--Virgo analysis. 

\begin{figure}[h!]
    \centering
    \includegraphics[width = 0.6\columnwidth, clip=true]{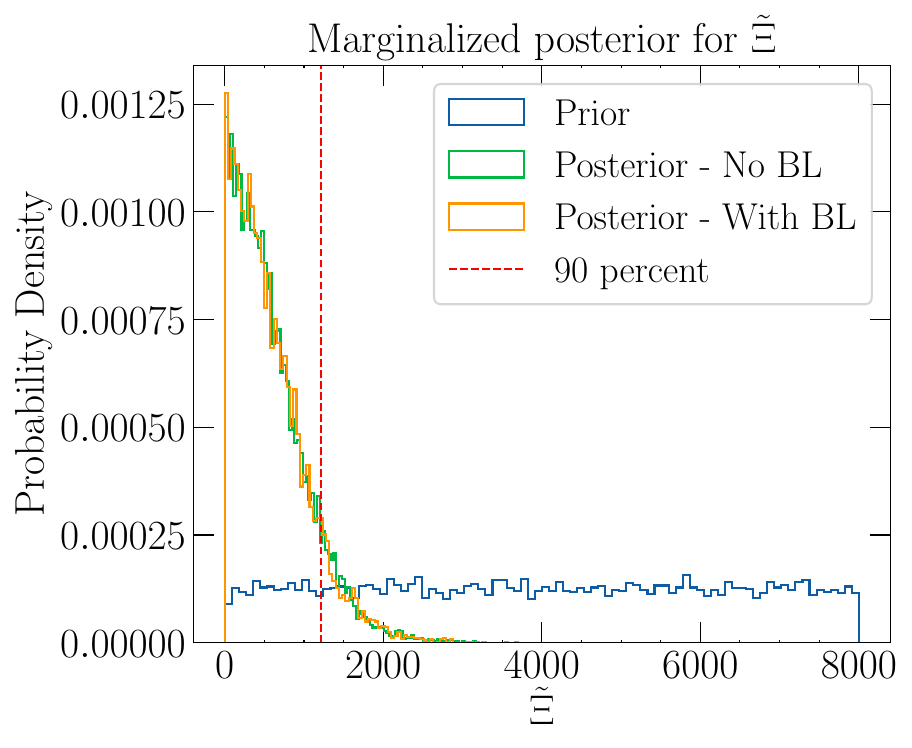}
    \caption{
    Marginalized posterior distribution for $\tilde{\Xi}$ for the GW170817 event with and without the marginalized binary Love relations~\cite{Yagi:2015pkc,Carson:2019rjx}, and the $90\%$ credible interval.
    The prior probability distribution for $\tilde{\Xi}$ is plotted in blue, which the marginalized posterior for $\tilde{\Xi}$ is plotted in green (corresponding to sampling in $\Lambda_{A,B}$) and orange (corresponding to sampling in $\Lambda_s$).
    The $90\%$ credible interval for the latter posterior is given by the vertical dashed line at $\tilde{\Xi}=1200$. 
    Observe that the GW170817 data is sufficiently informative to significantly constrain $\tilde{\Xi}$ as compared to the prior, while remaining independent of the use of the binary Love relations. 
    }
    \label{fig:marginalized-xibar-GW170817}
\end{figure}

Finally, we performed another consistency check of our result for the posterior probability distribution for $\tilde{\Xi}$, shown in figure~\ref{fig:marginalized-xibar-GW170817}, by considering different prior distributions for $\tilde{\Xi}$.
We used a uniform prior in $\Xi_{A/B}$ to obtain the prior on $\tilde{\Xi}$; we also sampled using a log-uniform prior on $\tilde{\Xi}$.
In both these cases, we find that when we divide the posterior probability distribution, multiply by our flat prior for $\tilde{\Xi}$, and renormalize, we obtain a distribution statistically consistent with that shown in figure~\ref{fig:marginalized-xibar-GW170817}.

\subsection{Estimating the significance of the tidal terms}
To estimate at what frequency/radius tidal effects start to appreciably affect the phase of emitted gravitational waves, and as another check on the robustness of our analysis, here we compare the leading order phase contributions of the adiabatic and dissipative tides to the leading phase of a Newtonian binary of point particles. 
In particular, we compute the fraction $p$ of the tidal phase with respect to the Newtonian phase for both the conservative and dissipative tidal effects. 
We take the fiducial value of $p = 0.025$. 
The leading order phase contribution for $\tilde{\Xi}$ is given in equation~\eqref{eq:dissipative-tidal-phase-contribution}; we recall that the leading order phasing contribution of $\tilde{\Lambda}$ is \cite{Flanagan:2007ix}
\begin{align}
    \Delta \Psi_{\Lambda}
    =
    -
    \frac{117}{256}\frac{\tilde{\Lambda}}{\eta_{\mathrm{sym}}} u^5
    ,
\end{align}
while the leading order phase contribution for a Newtonian binary of point particles is (for a review, e.g. \cite{Maggiore-vol-1})
\begin{align}
    \Psi_{pp}
    =
    \frac{3}{128}\frac{1}{\eta_{\mathrm{sym}}} u^{-5} 
    .
\end{align}
Setting $\Delta \Psi_{\Lambda}/\Psi_{pp} = p$, and solving for the gravitational wave frequency $f$ (or the orbital radius $r$, using the relation $u^2=GM/(rc^2)$), we obtain
\begin{align}
    f &\approx 467 \mathrm{Hz} \left( \dfrac{m}{2.8 M_{\odot}} \right)^{-1} \left( \dfrac{p}{0.025}\right)^{3/10} \left( \dfrac{\tilde{\Lambda}}{575}\right)^{-3/10}, \nonumber \\ 
    r &\approx 55 \mathrm{km} \left( \dfrac{m}{2.8 M_{\odot}} \right) \left( \dfrac{p}{0.025}\right)^{-1/5} \left( \dfrac{\tilde{\Lambda}}{575}\right)^{1/5}
    .
\end{align}
For the dissipative tidal effect, setting $\left|\Delta \Psi_{\tilde{\Xi}}/ \Psi_{pp}\right| = p$, we solve the resulting transcendental equation numerically for $f$ (or $r$),
\begin{align}
    -\dfrac{75}{32} \tilde{\Xi} u^8 \log u = p .
\end{align}
Setting $p=0.025$, $\tilde{\Xi}=1200$ (based on the constraint from GW170817), and $m=2.8 M_{\odot}$, we find $f \approx 253 \mathrm{Hz}$ and $r \approx 84 \mathrm{km} \sim 8 R_A$. 
In other words, dissipative tidal effects typically become important at an earlier stage of the inspiral than conservative tidal effects. 
Note that for a fixed fraction $p$, a larger (smaller) $\tilde{\Xi}$ will contribute significantly at a lower (higher) frequency--that is the tidal effects grow stronger as the two stars in the binary move faster.
If we set $\tilde{\Xi}=1200$, from $f \approx 253 \mathrm{Hz}$ until $f \approx 1569 \mathrm{Hz}$ (corresponding to the last stable orbit for $m=2.8 M_{\odot}$ mass binary), $\Delta \Psi_{\tilde{\Xi}}$ accumulates $1.86$ GW cycles (see equation~\eqref{eq:dissipative-tidal-phase-contribution}).
For reference, over the entire inspiral (from 10Hz to 1569Hz), $\Delta \Psi_{\tilde{\Xi}}$ contributes $2.5$ GW cycles, while $\Psi_{pp}$ contributes $\sim 6060$ GW cycles. 
In conclusion, most of the contribution from $\tilde{\Xi}$ is accumulated during the late stage of binary inspiral.

\subsection*{Data availability}

All relevant data that supports the findings of this study are available at \url{https://doi.org/10.5281/zenodo.11626502}.

\subsection*{Code availability}

The code we used to perform our analysis is available at \url{https://doi.org/10.5281/zenodo.11589416}.

\subsection*{Acknowledgements}
We would like to thank Jorge Noronha, Jaki Noronha-Hostler, Colm Talbot, Yumu Yang, Mauricio Hippert, and Mark Alford for helpful discussions.
We acknowledge support from the National Science Foundation through award PHY-2207650 (JLR, AH, NY), from the University of Illinois Graduate College Dissertation Completion Fellowship (RC).  
We also acknowledge the Illinois Campus Cluster Program, the Center for Astrophysical Surveys (CAPS), and the National Center for Supercomputing Applications (NCSA) for the computational resources that were used to produce the results of our paper. 
\subsection*{Author contribution statement}
J.L.R. , R.S.C. and A.H.K.R. performed the data analysis. 
All authors contributed to the writing of the manuscript, and to the interpretation of the results. 

\subsection*{Competing interests statement}
The authors declare no competing interests.

\subsection*{Corresponding authors}
\begin{enumerate}
\item Justin L. Ripley, \url{lloydripley@gmail.com}
\item Abhishek Hegade K. R. \url{ah30@illinois.edu}
\end{enumerate}


\bibliography{thebib}


\begin{table}[h!]
    \centering
    \begin{tabular}{ c | c | c | c }
     $\tilde{\Xi}_{\mathrm{inj}}$ & $\tilde{\Xi}^{(\mathrm{O4})}_{\mathrm{MAP}}$ & $\tilde{\Xi}^{(\mathrm{O5})}_{\mathrm{MAP}}$ & $\tilde{\Xi}^{(\mathrm{CE})}_{\mathrm{MAP}}$ \\ 
     \hline 
     20   & $89^{+522}_{-88} $ & $16^{+301}_{-16} $ & $ 11^{+57}_{-11}$ \\  
     400  & $197^{+615}_{-197}$ & $270^{+301}_{-270}$ & $380^{+116}_{-123}$\\  
     800  & $528^{+560}_{-528}$ & $584^{+451}_{-541}$ & $771^{+132}_{-132}$ \\  
     1200 & $803^{+704}_{-704}$ & $974^{+584}_{-584}$ & $1175^{+136}_{-145}$    
    \end{tabular}
    \caption{The maximum of the posterior (MAP) with the symmetric $90\%$ credible interval about the MAP for each detector network and injected value of $\tilde{\Xi}$ (corresponding to the marginalized posteriors in figure~\ref{fig:posterior-xibar-histogram}). 
    The credible intervals are not symmetric about the MAP when one limit of the interval reaches a boundary of the posterior. 
    We can translate a measurement on $\tilde{\Xi}$ to a measurement of the stars averaged viscosity using equation~\eqref{eq:relating-zeta-and-xibar}. These averaged viscosities are quoted in the abstract.}
    \label{fig:posterior-xibar-table}
\end{table}

\end{document}